\documentclass[10pt,twocolumn,superscriptaddress,showpacs,preprintnumbers,amsmath,amssymb,prl,aps]{revtex4}
\usepackage{graphicx}
\usepackage{amssymb,amsmath}
\usepackage{float}
\begin{document}

%\preprint{}

\title{First-Order Reorientation of the Flux-Line Lattice in CaAlSi}

\author{P.K. Biswas}
\email[]{P.K.Biswas@warwick.ac.uk}
\affiliation{Physics Department, University of Warwick, Coventry, CV4 7AL, United Kingdom}

\author{M.R. Lees}
\affiliation{Physics Department, University of Warwick, Coventry, CV4 7AL, United Kingdom}

\author{G. Balakrishnan}
\affiliation{Physics Department, University of Warwick, Coventry, CV4 7AL, United Kingdom}

\author{D.Q. Liao}
\affiliation{Physics Department, University of Warwick, Coventry, CV4 7AL, United Kingdom}

\author{D.S. Keeble}
\affiliation{Physics Department, University of Warwick, Coventry, CV4 7AL, United Kingdom} 

\author{J.L. Gavilano}
\affiliation{Paul Scherrer Institut, CH-5232 Villigen PSI, Switzerland}

\author{N. Egetenmeyer}
\affiliation{Paul Scherrer Institut, CH-5232 Villigen PSI, Switzerland}

\author{C.D. Dewhurst}
\affiliation{Institut Laue-Langevin, 6 Rue Jules Horowitz, F-38042 Grenoble, France}

\author{D.McK. Paul}
\affiliation{Physics Department, University of Warwick, Coventry, CV4 7AL, United Kingdom}

\date{\today}

\begin{abstract}
The flux line lattice in CaAlSi has been studied by small angle neutron scattering. A well defined hexagonal flux line lattice is seen just above $H_{c1}$ in an applied field of only 54~Oe. A $30^\circ$ reorientation of this vortex lattice has been observed in a very low field of 200~Oe. This reorientation transition appears to be of first-order and could be explained by non-local effects. The magnetic field dependence of the form factor is well described by a single penetration depth of $\lambda=1496(1)$~\AA~and a single coherence length of $\xi=307(1)$~\AA~at 2~K. At 1.5~K the penetration depth anisotropy is ${\gamma}_{\lambda}=2.7(1)$ with the field applied perpendicular to the $c$ axis and agrees with the coherence length anisotropy determined from critical field measurements.
\end{abstract}

%\keywords{}

\maketitle

The ternary silicide superconductor CaAlSi (CAS)~\cite{Imai2, Lorenz, Lorenz2, Sagayama} has the same AlB$_2$-type layered structure as MgB$_2$~\cite{Nagamatsu}. CaAlSi exhibits a number of interesting superconducting properties, the study of which can provide an insight into the factors leading to high superconducting transition temperatures $\left(T_c\right)$ in this class of materials. 

Neutron and x-ray diffraction studies have shown that there are two possible arrangements for the atoms in the AlSi layers in CaAlSi. These layers stack along the $c$ axis in a sequence $\left(AABBB\right)$ in five-fold $5H$-CAS and $\left(AAABBB\right)$ in six-fold $6H$-CAS. Further distortions produce either corrugated or flat AlSi layers within the multi-stack structures~\cite{Sagayama}. An unmodulated phase ($1H$-CAS) can be grown by controlled cooling from the molten state~\cite{Kuroiwa3}.

The superconducting properties of CAS, including $H_{c2}\left(T\right)$ and $T_c$ (5.7 to 7.8~K), change with modulation, as does the anisotropy in the superconducting parameters $\gamma_H=H^{ab}_{c2}/H^{c}_{c2}$ and $\gamma_\lambda=\lambda_c/\lambda_{ab}$, although $\gamma$ values of 2-3 indicate that these materials are only moderately anisotropic~\cite{Imai, Ghosh, Kuroiwa2}. The heat capacity of $6H$-CAS below $T_c$ is well explained within the framework of the BCS theory with strong-coupling, with a single superconducting gap, $2\Delta$, at $T=0$~K giving $2\Delta/k_{B}T_{c}= 4.07$~\cite{Lorenz2}. While the $T_c$ of $1H$-CAS decreases with applied pressure $P$, $dT_c/dP = +0.21$~K/GPa for $6H$-CAS~\cite{Lorenz2, Boeri}.

The electronic structure of CAS consists of $\sigma$ and $\pi$ bands derived from hybridized (Al,Si) $s$ and $p$ states and Ca $s$, $p$ and $d$ states~\cite{Shein,Huang,Mazin}. In $6H$-CAS calculations show that there are two disconnected cylindrical Fermi-surfaces which have two-dimensional character~\cite{Kuroiwa5}. Different measurement techniques suggest different scenarios for the superconducting gap. ARPES measurements indicate that in $6H$-CAS there are two superconducting gaps with equal magnitudes~\cite{Tsuda}. Muon spin rotation studies of the field dependence of penetration depth $\lambda$~\cite{Kuroiwa2} and optical measurements both suggest an anisotropic or multi-gapped structure~\cite{Lupi}. In contrast, tunnel-diode resonator measurements and break-junction tunneling spectroscopy both point to a single weakly anisotropic $s$-wave gap in $6H$-CAS~\cite{Prozorov, Kuroiwa4}. 

In this letter we report the results of a small angle neutron scattering (SANS) study of the magnetic flux line lattice (FLL) in the $6H$ phase of CAS. SANS is a powerful technique for studying the mixed state of type-II superconductors~\cite{Cubitt} and has often been used to investigate the symmetry of the pairing mechanism and the macroscopic physics of the FLL~\cite{DeWilde,Paul,Kogan,Gurevich,Nakai,Levett,Kogan2,Park}. We demonstrate that the FLL in CAS has a reorientation transition at a relatively small applied magnetic field, similar to that seen in MgB$_2$ at much higher fields~\cite{Cubitt2}. This low field transition is probably due to non-local effects rather than multi-band physics as suggested for MgB$_2$. Our measurement of the field dependent form factor from the field distribution is explained by a single coherence length, and the anisotropy of this coherence length is the same as the anisotropy of the penetration depth. Both features are very unlikely to occur in a multi-band superconductor.

\begin{figure*}[htb]
\begin{center}
\includegraphics[width=1.5\columnwidth]{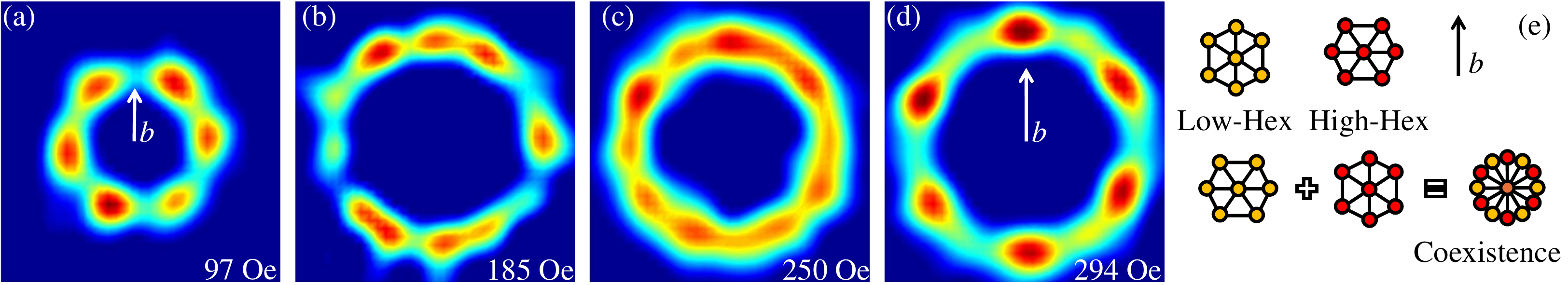}
\caption{\label{Figure1Biswas} (a-d) SANS diffraction patterns of CAS taken at 2~K in applied magnetic fields of 97, 185, 250, and 294~Oe respectively. (e) Schematic diagram of the FLL patterns in real-space (upper panel) and the corresponding diffraction patterns (lower panel).}
\end{center}
\end{figure*}

Single crystals of $6H$-CAS were prepared by the Bridgman method~\cite{superlattice}. Polycrystalline ingots of CAS were made by melting stoichiometric mixtures of calcium shot ($99.99\%$), aluminum shot ($99.999\%$) and silicon pieces ($99.99\%$) in an arc furnace under flowing argon gas. Each as cast ingot was placed in boron nitride crucible with a conical shaped bottom, and then sealed in a quartz tube under vacuum. The tube was placed in a vertical Bridgman furnace, heated to $1010^\circ$C at $100^\circ$C/h, and then held at this temperature for 24~h. Crystal growth was carried out by lowering the tube at a rate of 3~mm/h. 

SANS measurements were performed using the D22 instrument at the Institut Laue-Langevin, Grenoble, France. Additional anisotropy data were collected using the SANS I instrument at the Paul Scherrer Institut, Villigen, Switzerland ~\cite{SANSinstruments}. The samples were mounted with the $c$ axis parallel to the neutron beam direction to access the Bragg peaks. To scan through the Bragg condition for a diffraction spot, a rocking curve was performed by tilting or rotating the sample, cryostat and magnet together about a horizontal or vertical axis. For all measurements, the sample was cooled to base temperature in an applied magnetic field $H$, and the data collected while warming the sample in the same field. Background scattering was measured above $T_c$ and subtracted from the low-temperature data.

Data were collected in applied fields of 54 to 2000~Oe. Figs.~\ref{Figure1Biswas}(a)-(d) show typical diffraction patterns from the FLL of CAS measured at $2$~K in fields of (a) 97, (b) 185, (c) 250, and (d) 294~Oe applied parallel to the $c$ axis. The only previous SANS measurements on $6H$-CAS suggested that the FLL was not perfectly hexagonal~\cite{Kuroiwa}. In our measurements, a perfectly hexagonal lattice was found for all applied fields. We find a well defined FLL in an applied field of just 54~Oe. This is one of the lowest fields in which a FLL has been observed using the SANS technique. This field is also much smaller than the reported value of $H_{c1}$ for $6H$-CAS~\cite{Imai}. However, from magnetization versus field measurements, we estimate $H_{c1}=50$~Oe. Just above $H_{c1}$ the inter vortex distance is several times larger than the penetration depth and any inhomogeneity may be expected to disrupt the FLL leading to disordered vortex clusters. 

In low fields [Fig.~\ref{Figure1Biswas}(a)] the Bragg peaks in the diffraction pattern,denoted here as Low-Hex, appear at $30^\circ$ to the $b$ axis of the crystal. With increasing field, a second hexagonal diffraction pattern appears oriented along the $b$ axis [see Figs.~\ref{Figure1Biswas}(b) and~\ref{Figure1Biswas}(c)]. This means that the FLL has now formed two domains with an angular separation of $30^\circ$. As the applied field is increased further the FLL transforms into a single domain with Bragg peaks oriented along the $b$ axis and referred to as High-Hex [Fig.~\ref{Figure1Biswas}(d)]. We do not observe any intermediate structures or any continuous change in the positions of the diffraction peaks during the reorientation process. These observations suggest that the transition between the High and Low-Hex phases is first-order in character. No further reorientations of the FLL were observed in applied fields of up to 2~kOe. An earlier SANS study of $6H$-CAS found no evidence for a FLL reorientation as the measurements were not carried out in sufficiently low applied fields~\cite{Kuroiwa}. 

\begin{figure}[tb]
\begin{center}
\includegraphics[width=1.0\columnwidth]{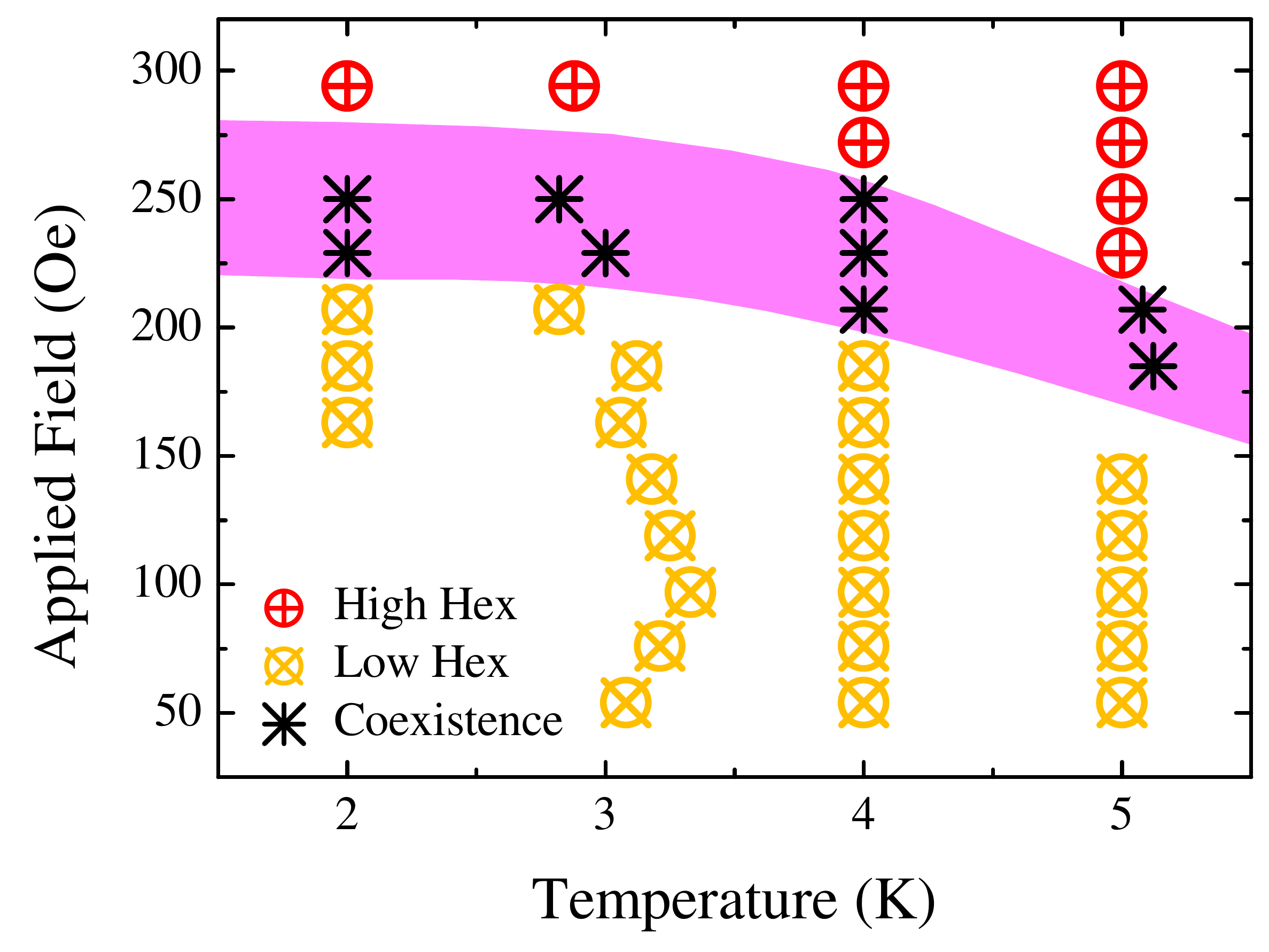}
\caption{\label{Figure2Biswas} (Color online) $H$-$T$ phase diagram of CAS indicating the temperatures and applied fields at which we observe either a High-Hex or a Low-Hex state for the FLL. A shaded region in which the two states coexist is also marked.}
\end{center}
\end{figure}

\begin{figure}[tb]
\begin{center}
\includegraphics[width=1.0\columnwidth]{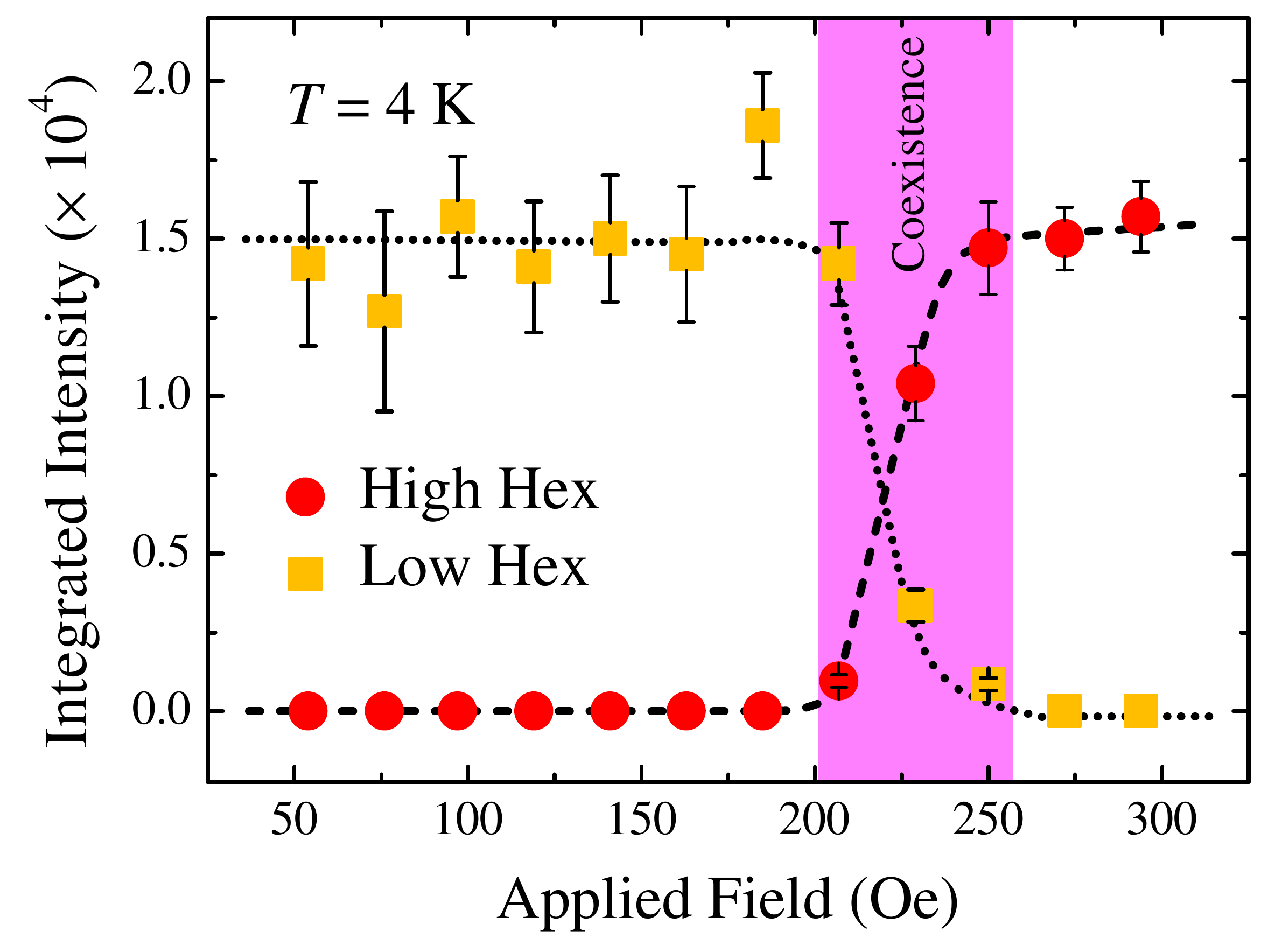}
\caption{\label{Figure3Biswas} (Color online) Standard monitor normalized intensity of the Bragg peaks for the High-Hex and Low-Hex phases of CAS. The dotted and dashed lines are guides to the eye. The shading indicates the region in which the two FLL phases coexist.}
\end{center}
\end{figure}

An $H$-$T$ phase diagram of CAS is shown in Fig.~\ref{Figure2Biswas} indicating the regions in which we observe either a purely Low-Hex or a High-Hex phase separated by a shaded region in which the two FLL structures coexist. Fig.~\ref{Figure3Biswas} shows the variation of the integrated intensity of the Bragg spots for the High and Low-Hex states with applied magnetic field at 4~K. A sudden change of intensity for the two states occurs through the narrow window of coexistence. In addition to the first-order nature of the transition, the coexistence of the two phases is due to a combination of shape demagnetization effects and pinning in the sample.

What drives the FLL reorientation in CAS? In MgB$_2$, a $30^\circ$ reorientation in the FLL has been associated with the suppression of the smaller of the two superconducting gaps present in the material. However, the reorientation field $H_r$, for MgB$_2$ is over 5~kOe $\left(H_{c2}/H_r\approx5\right)$ and the reorientation process is second-order~\cite{Cubitt2}. The $30^\circ$ reorientation of the FLL reported here occurs in a field of only 200~Oe which is a small fraction of $H_{c2}=8$~kOe for this material, $\left(H_{c2}/H_r\approx40\right)$. For $6$-H CAS, the experimental data suggest that if two gaps are present the ratio of their magnitudes $\Delta_{c1}/\Delta_{c2}\leq3.5$~\cite{Tsuda, Lupi}. We conclude, therefore, that the reorientation of the FLL in CAS is not caused by the closing of the smaller of two superconducting gaps.

In some ways the FLL transition in CAS more closely resembles the (apparently) first-order 45$^{\circ}$ reorientation between two rhombic FLL phases observed in Lu and Y borocarbide~\cite{Dewhurst,Paul,Vinnikov}. $H_r$ is 250 and 1500~Oe for Lu and Y borocarbide respectively with $\left(H_{c2}/H_r\approx50\right)$ Changes in the symmetry of the FLL and its orientation with respect to the high-symmetry directions in the crystal result from an anisotropy in the penetration depth and can be understood by considering non-local corrections to the London model~\cite{Yethiraj,Dewhurst,Paul,Kogan3}. When the inter-vortex distance is comparable to the penetration depth, the non-local effects play their part to align the vortices with the strong vortex-vortex interactions.

Non-local effects are expected to be significant in a low-$\kappa$ superconductor like CAS. At the reorientation field $H_r=200$~Oe, the inter-vortex distance in the hexagonal FLL is the same order of magnitude as the penetration depth. In addition, the morphology of the FLL and the $30^\circ$ reorientation reflect the underlying symmetry of the crystallographic lattice. These facts strongly suggest that non-local effects are driving the low-field reorientation of the FLL in CAS. To unambiguously establish this will require a calculation of the non-local corrections to a London model for a material with six-fold symmetry. This will be difficult, given the need for complicated Fermi surface averages.

\begin{figure}[tb]
\begin{center}
\includegraphics[width=1.0\columnwidth]{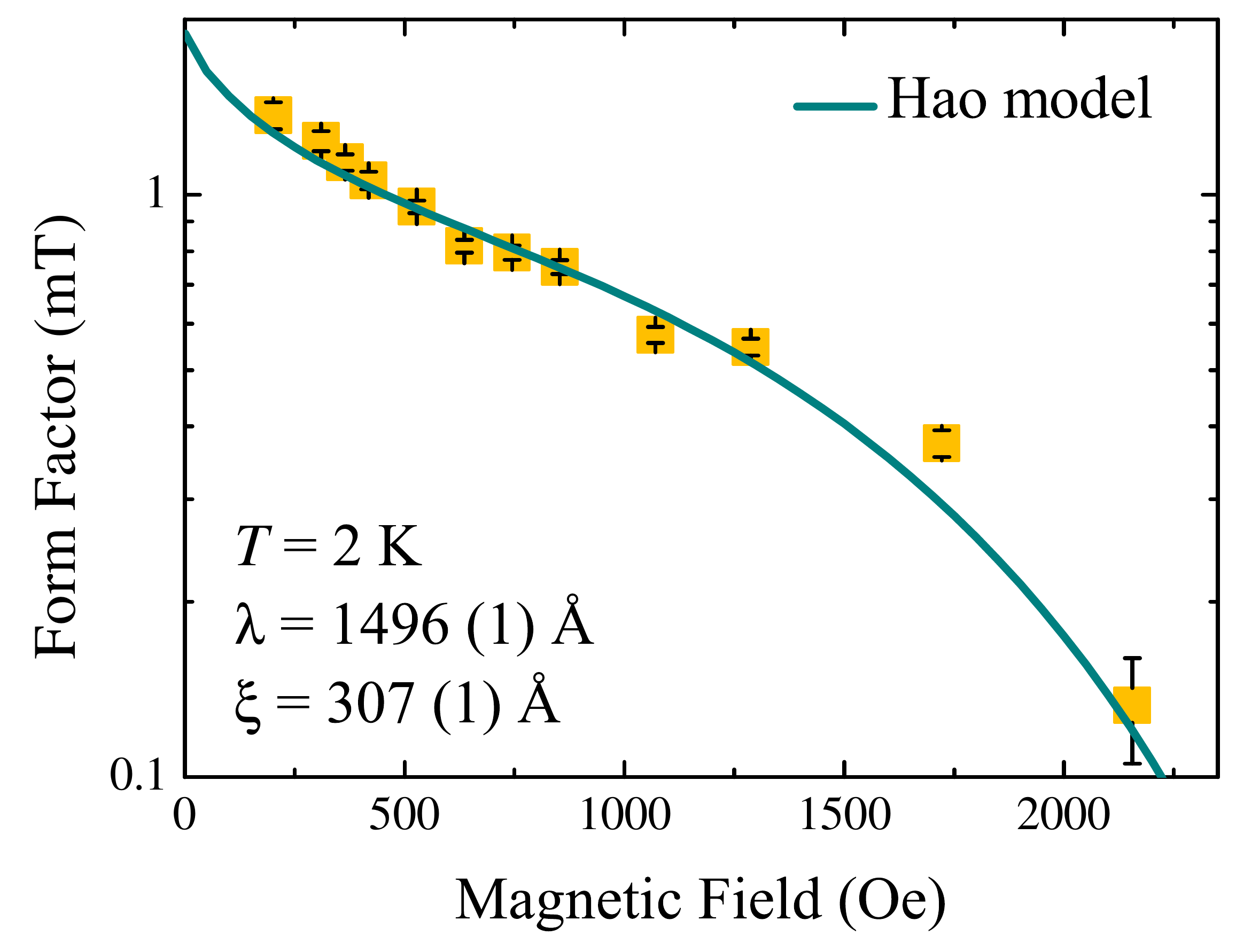}
\caption{\label{Figure4Biswas} (Color online) Form factor $F$ of CAS plotted on a log scale. The solid line is a fit to the data using the Hao model described in the text and in Ref.~\cite{Hao}.}
\end{center}
\end{figure}

Fig.~\ref{Figure4Biswas} shows the form factor $F$ at 2~K, extracted from the integrated intensity of the Bragg spots forming the FLL in CAS. The form factor provides a measure of the amplitude of the field modulation inside a type-II superconductor due to the formation of a FLL~\cite{Cubitt2}. According to the London model, for a conventional single band superconductor with a penetration depth and a coherence length that are independent of field, $F$ decreases exponentially with field~\cite{Yaouanc}. However, for an anisotropic superconductor, an expression for $F$ has been calculated by Hao \textit{et al.}~\cite{Hao} within the Ginzburg-Landau (GL) approximation.
\begin{equation}
F=\frac{3^{1/4}}{2\pi\sqrt{2}}\frac{\sqrt{{\Phi_{0}}B}f^{2}\xi_{v}}{\lambda^{2}}K_{1}\left(\frac{2\pi\sqrt{2}}{3^{1/4}}\xi_{v}\sqrt{B/\Phi_{0}}\right)
\label{Haoeq}
\end{equation}
with
\begin{subequations}\label{subHaoeq}
\begin{align}
\xi_{v}&=\xi\left(\sqrt{2}-\frac{0.75}{\kappa}\right)\sqrt{\left(1+b^{4}\right)\left[{1-2b{(1-b)}^{2}}\right]},\label{subHao1}\\
f^2&=1-b^{4}, \label{subHao2}
\end{align}
\end{subequations}
$K_n(x)$ is a modified Bessel function of $n^{\rm{th}}$ order, $\Phi_{0}= 2.068 \times10^{-15}$~Wb is the magnetic flux quantum, $\kappa=\lambda/\xi$ is the GL parameter, $B_{c2}=\Phi_{0}/(2\pi{\xi}^{2})$ is the upper critical field, and $B=bB_{c2}$ is the applied field~\cite{Brandt, Yaouanc}. The fit yields $\lambda=1496(1)$~\AA, $\xi=307(1)$~\AA, and $\kappa=4.9(1)$. This $\kappa$ is consistent with the value of 5.2 reported by Imai \textit{et al}.~\cite{Imai}. $\xi$ is $50\%$ larger than the value extracted from $H_{c2}$ measurements on the same sample. In a study of MgB$_2$, the increase in $F$ at low field was attributed to a change in the superfluid density~\cite{Cubitt2}. As shown here, such a conclusion is not required for CAS.

\begin{figure}[tb]
\begin{center}
\includegraphics[width=1.0\columnwidth]{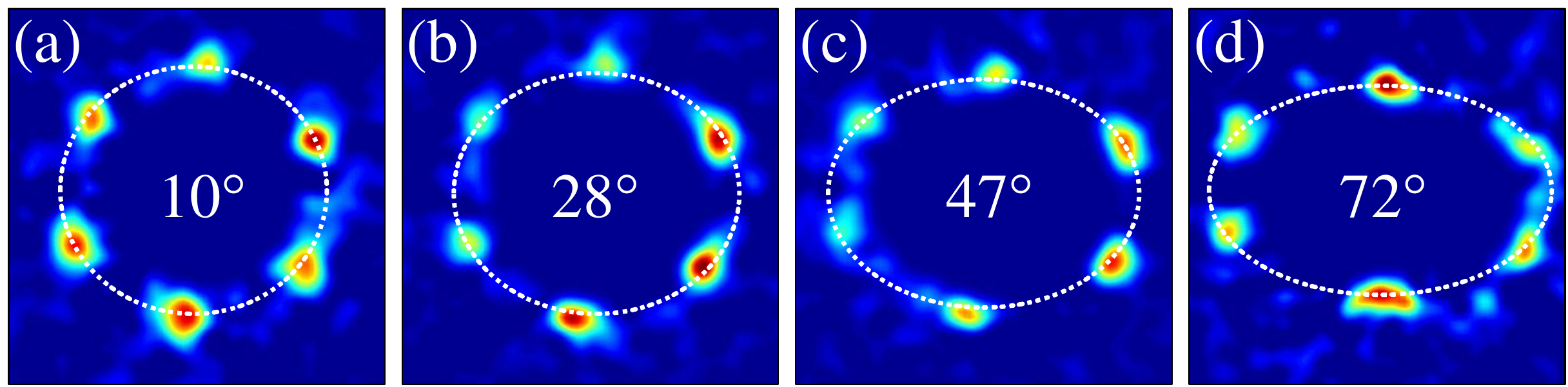}
\caption{\label{Figure5Biswas} (Color online) (a-d) Anisotropic FLL of CAS at $1.5$~K and a field of 3~kOe applied at $10^\circ$, $28^\circ$, $47^\circ$, and $72^\circ$ respectively to the $c$~axis of the crystal.}
\end{center}
\end{figure}

The penetration depth anisotropy $\gamma_{\lambda}$, can be extracted by rotating the applied magnetic field away from the $c$~axis and measuring the ratio of the major to minor axes of the ellipse ($\epsilon$) connecting the Bragg peaks. Figs.~\ref{Figure5Biswas}(a)-(d) show the diffraction patterns of CAS taken at 1.5~K in a field of 3 kOe applied at $10^\circ$, $28^\circ$, $47^\circ$, and $72^\circ$ respectively to the $c$ axis. As the angle between the applied field and the $c$~axis increases, the diffraction pattern is distorted towards an elliptical shape, since the screening currents circulating around a vortex must cross the basal plane. Campbell \textit{et al.}~\cite{Campbell} studied the structure of a vortex lattice in anisotropic, uniaxial superconductors, for magnetic fields applied at an angle $\psi$ to the principal axis. According to their model based on the London approach, $\epsilon$ is related to ${\gamma}_{\lambda}$ in the following way,

\begin{equation}
{\epsilon}^2=\frac{\gamma^{2}_{\lambda}}{{\sin}^2\psi+\gamma^{2}_{\lambda}{\cos}^2\psi}.
\label{Campbell}
\end{equation}

Fig.~\ref{Figure5Biswas} shows the variation of $\epsilon$ as a function of $\psi$ for CAS measured at 1.5~K in a field of 3~kOe. A fit to the data using Eq.~\ref{Campbell} is indicated by the solid line yielding an anisotropy, ${\gamma}_{\lambda}=2.7(1)$. The value of ${\gamma}_{\lambda}$ is in excellent agreement with previous values of ${\gamma}_{\xi}$ determined from magnetic and transport measurements~\cite{Imai,Ghosh} and slightly larger than the value of 2 obtained by Kuroiwa \textit{et al.}~\cite{Kuroiwa} from SANS measurements.

Close to $T_c$ the anisotropic GL equations for a clean superconductor with an arbitrary gap anisotropy yield ${\gamma}_{\lambda}={\gamma}_{\xi}$. At lower $T$, however, these two quantities may both depend on $T$ and are not necessarily the same. For example, in the case of MgB$_2$ calculations for a weakly coupled two-band anisotropic superconductor showed that ${\gamma}_{\lambda}\left(T\right)$ and ${\gamma}_{\xi}\left(T\right)$ are an increasing and decreasing function of $T$ respectively~\cite{Kogan4, Miranovic}. In CAS the equality of ${\gamma}_{\lambda}$ and ${\gamma}_{\xi}$ at 1.5~K may reflect the fact that the morphology of FLL is established at higher $T$ and which then gets pinned as $T$ is reduced. Alternatively it may be indicative of a more isotropic character for the Fermi surface in this material.   

\begin{figure}[tb]
\begin{center}
\includegraphics[width=1.0\columnwidth]{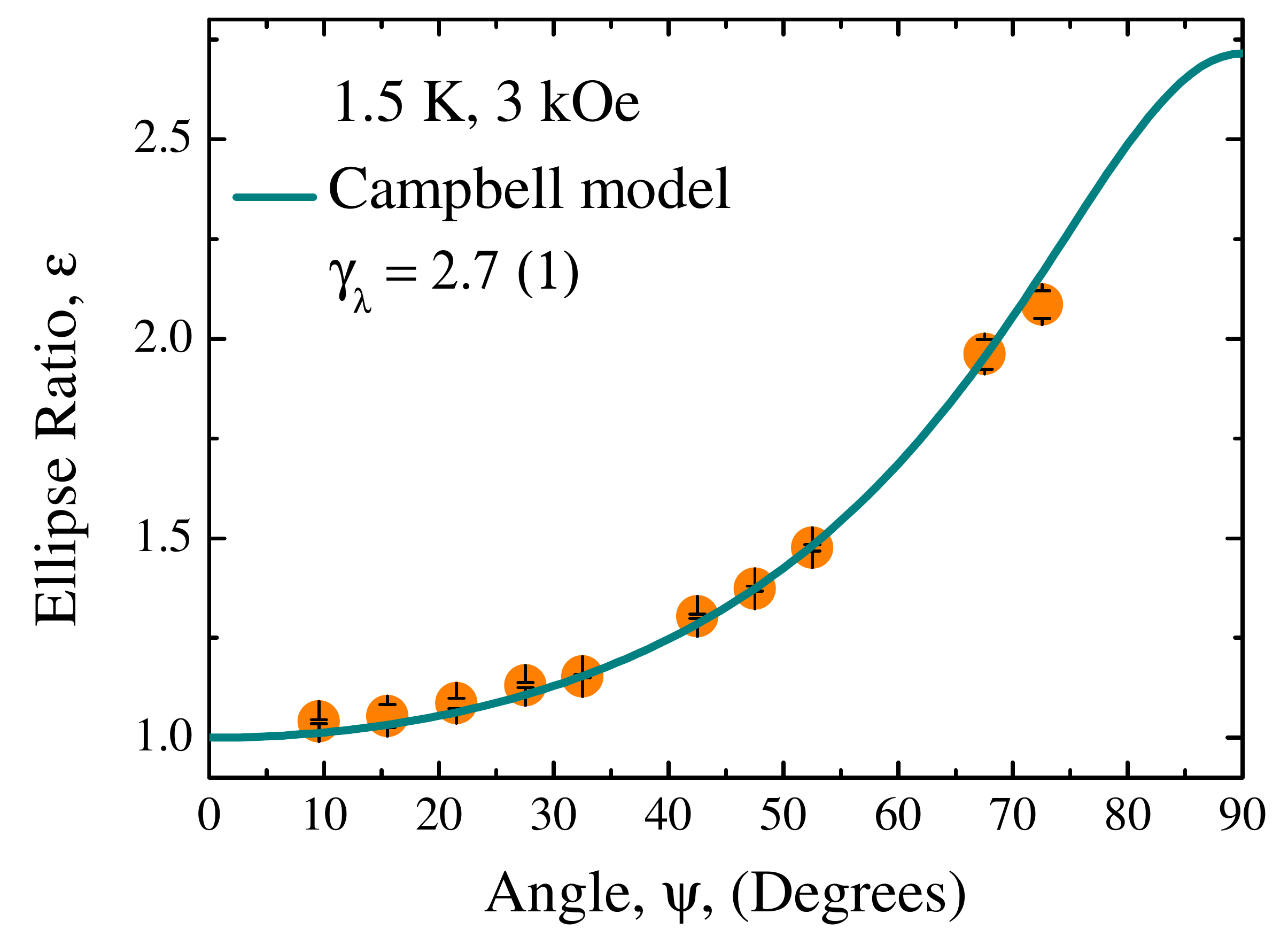}
\caption{\label{Figure6Biswas} (Color online) ${\gamma}_{\lambda}$ as a function of angle, $\psi$ at 1.5~K and applied field of 3 kOe for CAS. The solid line is  fit to the data using the Campbell model and yields ${\gamma}_{\lambda}=2.7(1)$ at $\psi=90^\circ$.}
\end{center}
\end{figure}

This work was supported by the Engineering and Physical Sciences Research Council (EPSRC). Some of the equipment used in this research was obtained through the Science City Advanced Materials project: Creating and Characterizing Next Generation Advanced Materials project, with support from Advantage West Midlands (AWM) and part funded by the European Regional Development Fund (ERDF). DSK thanks the Science City Research Alliance and the HEFCE Strategic Development Fund for financial support. PKB would like to thank the Midlands Physics Alliance Graduate School (MPAGS) forfinancial support. The authors would like to thank V. Kogan for valuable discussions and interest in this work.



\begin{thebibliography}{41}
\bibitem[{SAN()}]{SANSinstruments}
\bibinfo{note}{D22 was configured in a high resolution mode with a mean
  wavelength of $14$~\AA~and a wavelength spread of $10$\%, collimation of
  $17.6$~m, and an area detector at a distance of $17.6$~m to the sample. An
  additional circular aperture of diameter $20$~mm was also used before the
  sample to better define the scattered beam at the detector and to equalize
  the horizontal and vertical divergences. At SANS I, incident neutrons of
  various wavelengths between $5$-$10$~\AA\ were selected with a wavelength
  spread of $10$\% and collimated over a distance of 8 to 18~m before the
  sample. Diffracted neutrons were collected with a position-sensitive
  two-dimensional multidetector located $13$-$20$~m after the sample.}
\end{thebibliography}
\end{document}